\documentclass[twocolumn]{aastex701}
\graphicspath{{./}{figures/}}

\usepackage{xspace,subcaption}

\newcounter{subsubsubsectionctr}[subsubsection]

\renewcommand{\thesubsubsubsectionctr}{\thesubsubsection.\arabic{subsubsubsectionctr}}

\newcommand{\subsubsubsection}[1]{%
    \refstepcounter{subsubsubsectionctr}%
    \vspace{0.5\baselineskip}%
    \begin{center}{\normalsize\textit{\thesubsubsubsectionctr.\space#1}}\end{center}\vspace{-0.4\baselineskip}}

\begin{document} 

\title{Exploring a cosmic ray inverse-Compton origin to the SZ-to-X-ray pressure deficit in the cool core cluster ZwCl 3146}

\author[0000-0002-1616-5649]{Emily M. Silich}
\affiliation{Cahill Center for Astronomy and Astrophysics, California Institute of Technology, Pasadena, CA 91125, USA}
\email[show]{esilich@caltech.edu}  

\author[0000-0002-8213-3784]{Jack Sayers}
\affiliation{Cahill Center for Astronomy and Astrophysics, California Institute of Technology, Pasadena, CA 91125, USA}
\email{}

\author[0000-0003-3729-1684]{Philip F. Hopkins}
\affiliation{Cahill Center for Astronomy and Astrophysics, California Institute of Technology, Pasadena, CA 91125, USA}
\email{}

\author[0000-0001-5725-0359]{Charles Romero}
\affiliation{Department of Physics and Astronomy, University of Pennsylvania, 209 South 33rd Street, Philadelphia, PA, 19104, USA}
\email{}

\author[0000-0002-8472-836X]{Brian Mason}
\affiliation{National Radio Astronomy Observatory, 520 Edgemont Rd, Charlottesville, VA 22903}
\email{}

\author[0000-0003-1842-8104]{John Orlowski-Scherer}
\affiliation{Department of Physics and Astronomy, University of Pennsylvania, 209 South 33rd Street, Philadelphia, PA, 19104, USA}
\email{}

\author[0000-0003-0167-0981]{Craig L. Sarazin}
\affiliation{Department of Astronomy and Virginia Institute for Theoretical Astronomy, University of Virginia, P.O. Box 400325, Charlottesville, VA 22904, USA}
\email{}

\begin{abstract}
We explore the possibility that inverse-Compton (IC) scattering of cosmic microwave background photons by $\sim$GeV cosmic rays (CRs) injected by the central active galactic nucleus (AGN) in cool core (CC) clusters produces a non-negligible continuum-like X-ray signal that is easily misinterpreted as intracluster medium (ICM) thermal bremsstrahlung continuum. This is particularly relevant to the cooling flow problem--the lack of star formation relative to X-ray-inferred ICM cooling rates. Using ZwCl 3146, a relaxed CC system at $z=0.291$, we compare pressure profiles derived via X-rays and the thermal Sunyaev–Zel'dovich (SZ) effect. While SZ measurements probe only thermal ICM electrons, additional CR–IC emission would appear to boost the X-ray–inferred pressure. Relative to unity, we measure a $25\%$ decrement in $P_{SZ}/P_X$ within $100$~kpc of the ZwCl 3146 center at a statistical significance $\simeq3\sigma$, consistent with predicted deficits from CR–IC contamination in reasonable models of central AGN-driven CR injection. X-ray spectral fits of a two-component model with thermal ICM and CR-IC emission are consistent with CR-IC as the cause of this deficit. We test alternative explanations and systematics that could drive such a decrement, with leading-order systematics associated with halo triaxiality and AGN-driven cavities. On average, these systematics could bias $P_{SZ}/P_X$ low by $\simeq5\%$, a factor of two lower than our measurement uncertainty, with an object-to-object variation $\simeq5\%$. While our results establish that non-negligible CR-IC emission is plausible in ZwCl 3146, we stress that detailed studies of larger cluster samples are required to robustly assess whether CR-IC is relevant to the cooling flow problem.

\end{abstract}


\section{Introduction} 
Cool core (CC) galaxy clusters are characterized by central ($r \lesssim100$~kpc), rapidly-cooling regions of the intracluster medium (ICM) that exhibit radiative cooling timescales shorter than a few Gyr \citep[see][for a review]{Fabian1994}. The cooling rates associated with these CCs inferred from X-ray observations are significantly higher than observed star formation rates in the central brightest cluster galaxies (BCGs) or cold gas reservoir quantities allow \citep[e.g.,][]{Bregman2006,ODea2008,Peterson2001,McDonald2011,McDonald2018}. Historically, this discrepancy has been termed the ``cooling flow problem'', though modern studies have proposed a resolution in the form of a self-regulating cycle comprising the rapidly-cooling ICM as it falls onto the central BCG, which itself hosts an active galactic nucleus (AGN) that, fueled by the infalling gas, provides a form of mechanical heating that re-deposits energy into the surrounding medium and prevents catastrophic cooling of the ICM within the CC \citep[for a review, see][]{McNamara2007}. While these scenarios require a relatively fine tuning between the ICM thermodynamics and accretion rate onto the AGN, plausible models for this fine tuning have been proposed \citep[e.g.,][]{Voit2015}.

In support of such a self-regulating scenario, CCs tend to host strong radio AGNs \citep{Mittal2009,Sun2009}. The leptonic and kinetic power injected by radio jets in CCs is comparable to the apparent X-ray cooling luminosity within the CC \citep{Birzan2004,OSullivan2011,HlavacekLarrondo2012}, and AGN radio luminosities are weakly correlated with the X-ray CC size \citep{Liu2024}. However, the physics governing the regulation of ICM cooling and heating in CCs are not well established. The balance of ICM cooling with heating via feedback in galaxy groups and clusters is only self-regulated well on long timescales \citep{McDonald2018}, and while many AGN-driven heating mechanisms in CCs have been proposed, it is unclear to what extent these processes are responsible for the thermalization of energy in the ICM: for example, the dissipation of sound waves and weak shocks driven by expanding AGN bubbles \citep[e.g.,][]{Fabian2003,Ruszkowski2004,Forman2005,Mathews2006}, AGN-bubble-driven generation and dissipation of gravity waves via turbulence \citep{Zhuravleva2014,Zhuravleva2016,Li2020}, mixing of hot AGN bubble plasma with the ICM \citep{Yang2016,Hillel2016,Hillel2017}, and cosmic ray (CR) heating via Alfv\'en waves \citep{Pfrommer2013}. 

The correlation of radio AGN luminosity with X-ray CC luminosity implies that a significant population of CRs is being accelerated in CCs. The radio emission is primarily generated by high-energy leptons ($E \gg 10-100$ GeV), which, having short lifetimes ($\lesssim 10^{7}$ yr; see \citealt{Ruszkowski2023} for a review), typically generate emission only within a few $\sim$kpc of the CC centers \citep[e.g.,][]{deGasperin2012}. Most of the total CR lepton energy should, however, be contained by a population of low-energy ($E \sim 0.1-1$ GeV) leptons \citep{Hopkins2025CRs.CC}. In contrast to the high-energy CR leptons generating the radio emission, these low-energy leptons have much longer lifetimes ($\sim$ Gyr), so they should diffuse or stream out to larger radii ($\sim 100$ kpc) before losing most of their energy, comprising extended, ancient cosmic ray halos \citep[ACRHs;][]{Hopkins2025CRs.CC}. These ACRHs would produce $\sim$ keV X-ray emission via inverse Compton (IC) scattering with cosmic microwave background (CMB) photons. Since this CR population is peaked around one GeV, CR-IC from these halos would exhibit thermal continuum-like X-ray spectra \citep{Hopkins2025CRs.gal}. 

In massive clusters, X-ray emission from CR-IC would be insignificant relative to the thermal bremsstrahlung emission at large radii, but it would appear as $\sim$\,keV emission in the cluster cores at radii $\lesssim 100\,$kpc \citep{Hopkins2025CRs.CC}. Because the $0.1-1\,$GeV CR lifetime is of order the age of a cluster ($\sim$ Gyr), ACRHs should be present in a significant fraction of clusters. In addition, given the similarity of the power injected by radio jets in CCs and the apparent X-ray cooling luminosity, CR-IC from ACRHs could non-negligibly bias X-ray-inferred CC properties. The expected emissivity profile for CR-IC emission in these ACRHs is similar to the observed X-ray emission profiles in CCs \citep{Hopkins2025CRs.CC}. Therefore, interpreting a non-negligable fraction of the X-ray emission as CR-IC could alleviate the most challenging aspects of the cooling flow problem: the CR-IC contamination could explain why the apparent X-ray cooling rates of CCs are so large compared to constraints on observed cooling rates; the radio AGN properties could be so well-correlated with the apparent X-ray cooling luminosity because the AGN are the direct source of the low-energy leptons producing the CR-IC in X-rays, and the radio emission is tracing a younger population of high-energy leptons; and it would explain why CC clusters appear to follow universal profile shapes in their X-ray-inferred thermodynamic quantities \citep{Hopkins2025CRs.CC}.

Because the CR-IC spectra exhibit shapes similar to thermal continuum, observationally determining the fraction of CR-IC in X-rays via spectroscopy is extremely challenging. Perhaps the cleanest test for such CR-IC contributions involves comparison with the thermal Sunyaev-Zel'dovich (SZ) effect, which is the IC scattering of $\sim$keV thermal electrons in the ICM with CMB photons \citep[see][for a review]{Mroczkowski2019}. Since the number of thermal electrons in the ICM is many orders-of-magnitude larger than the number of CR leptons, and the thermal SZ effect is sensitive to the non-relativistic electron pressure, SZ measurements provide a robust measurement of the true thermal gas pressure of the CC. Conversely, the X-ray-derived pressure profiles of CCs are primarily driven by the X-ray-inferred density, which is derived from the observed soft X-ray emission. Therefore, the pressure derived from X-rays would in general be different, and most often boosted, relative to the true thermal gas pressure when CR-IC contributes to the X-ray continuum, and equal to the true thermal gas pressure in the absence of CR-IC contamination. So, the ratio of SZ-to-X-ray derived pressure ($P_{SZ} / P_{X}$) as a function of radius in CCs provides a potentially powerful test of these ACRH models, which we will leverage in this study. 

ZwCl 3146 is a massive \citep[$M_{500} \simeq 7.7 \times 10^{14}$ M$_{\odot}$;][]{Romero2020}, relaxed CC cluster \citep[e.g.,][]{Kausch2007} at $z = 0.291$ \citep{Allen1992}. Notably, ZwCl 3146 exhibits a large X-ray inferred cooling rate \citep[$\sim1000$ M$_{\odot}$ yr$^{-1}$][]{Edge1994,Egami2006,Kausch2007,McDonald2018}, and it also hosts a central radio source embedded within a diffuse radio minihalo extending out to $\sim 90$~kpc in radius \citep{Giacintucci2014}. Given the need to resolve the inner $\simeq100$~kpc of a CC to test for the presence of CR-IC contamination via pressure profile comparisons, deep, high-resolution X-ray and SZ data are required. ZwCl 3146 represents the only such CC with not only adequate available observations in each waveband (having 86 ks of archival \textit{Chandra} data and being detected at $61 \sigma$ significance with MUSTANG-2; \citealt{Romero2020}), but also a sufficiently low mm-wave-brightness central AGN 
so as to be disentangled from SZ signal \citep{Romero2020}. In this work, we explore the possibility that CR-IC generated by CRs injected by the central AGN are contributing significantly to the X-ray continuum in ZwCl 3146, and thus test whether such CR-IC could be biasing X-ray-inferred thermodynamic properties of CCs. 

In Section \ref{sec:data_analysis}, we describe the ZwCl 3146 SZ and X-ray data analysis, and Section \ref{sec:pressuregen} provides an overview of the calculation of the SZ- and X-ray-derived pressure profiles from these datasets. We compare these ZwCl 3146 pressure profiles in Section \ref{sec:results}, and highlight a detection of an SZ-to-X-ray pressure deficit in the central $\sim100$ kpc of ZwCl 3146. In Section \ref{sec:discussion}, we discuss physical scenarios that could contribute to the observed pressure deficit, including CR-IC from an ACRH as well as additional possible contaminants and systematics. Our conclusions are given in Section \ref{sec:conclusions}. Throughout this work, we assume a concordance cosmology: $H_0 = 70$ km s$^{-1}$ Mpc$^{-1}$, $\Omega_{m, 0} = 0.3$, unless otherwise specified. 

\section{Data Analysis} \label{sec:data_analysis}
\begin{figure*}[th!]
    \centering 
        \includegraphics[width=1\textwidth]{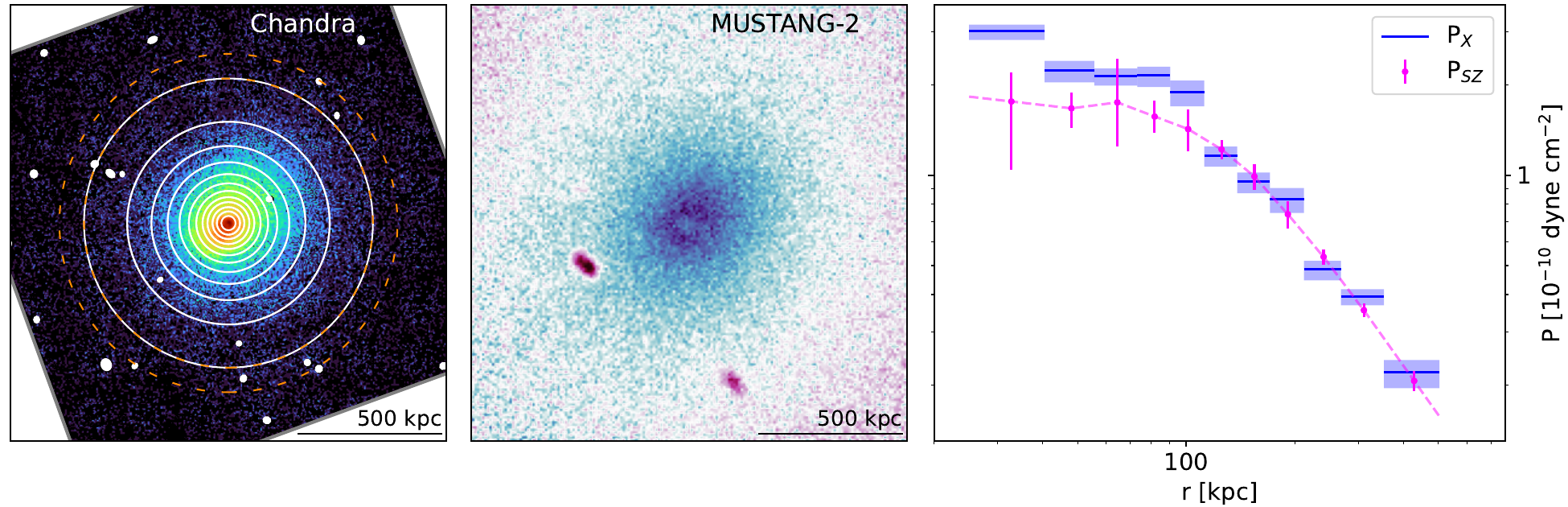} 
        \caption{\textit{Left:} $0.5-7$ keV (background-subtracted) \textit{Chandra} X-ray counts map of ZwCl 3146 with annular bins defining pressure profile extraction regions. \textit{Middle:} MUSTANG-2 SZ map of ZwCl 3146 for visualization. Note that the pressure profile is not derived from this map, but rather the time-ordered data. \textit{Right:} SZ- and X-ray-derived pressure profiles with $P_X$ normalized to $P_{SZ}$ using the radial bins outside of 100 kpc. Uncertainties are $1\sigma$.} \label{fig:pressures}
\end{figure*} 
 
\subsection{SZ} \label{subsec:SZdata}
We use MUSTANG-2 data which has been presented in earlier works, \citet{Romero2020,Romero2023}. Operating on the 100-m Green Bank Telescope (GBT), MUSTANG-2 achieves $10^{\prime\prime}$ resolution at full-width half-maximum (FWHM) with an instantaneous field of view (FOV) of $4^{\prime}.2$ \citep{Dicker2014a}. MUSTANG-2 observes targets via on-the-fly mapping. In the case of ZwCl 3146, Lissajous daisy scans, primarily of $2^{\prime}.5$ and $3^{\prime}.0$ scanning radii, were used.

In \citet{Romero2020}, an analysis of the MUSTANG-2 data in the time-domain was found to better recover the pressure profile at large radii relative to a more canonical map-domain analysis. The time-domain analysis relies on characterizing the noise in Fourier space. To create a map of MUSTANG-2 for visualization and comparison with the X-ray images, we consider each pixel to be a model component (see Sec. \ref{subsec:SZpressure}). The complexity here negates the potential for an explicit solution, and rather a (preconditioned) conjugate gradient descent is used (see again \citealt{Romero2020}). We show the resultant map in Figure~\ref{fig:pressures}, though, to reiterate, the pressure profile is not derived from this map.

\subsection{X-ray} \label{subsec:X-raydata}
We performed the \textit{Chandra} X-ray data reduction with \texttt{CIAO} version 4.16 \citep{ciao}, utilizing a modified version of the data reduction pipeline outlined in \citet{Silich2024} applied to two ZwCl 3146 datasets: \textit{Chandra} ObsIDs 909 and 9371, which are contained in the Chandra Data Collection ~\dataset[DOI: 10.25574/cdc.546]{https://doi.org/10.25574/cdc.546}. We calibrated the raw ACIS-I data for each ObsID using the CalDB version 4.11.3 with \texttt{chandra$\_$repro}. For each dataset, we performed an exposure correction with \texttt{fluximage}, point source identification via \texttt{wavdetect} \citep{Freeman2002}, and exclusion of point sources from the exposure-corrected data. We filtered the light curves for each observation with \texttt{deflare} and applied the identified good-time-intervals (GTI) to the data. After filtering, the total GTI is 86 ks between ObsIDs 909 and 9371. From these filtered datasets, we generated a ``clean'' (filtered, exposure-corrected, point-source-free) merged $0.5$--$7$ keV surface brightness ($S_X$) map for each ObsID with \texttt{flux\_obs}. 

We then constructed blank-sky background event files that are reprojected and scaled to match each of the ZwCl 3146 observations with \texttt{blanksky}. These CalDB blank-sky datasets, which are constructed by averaging deep (high-statistics), point source-free observations across large regions of the sky, characterize contributions from the \textit{Chandra} particle-induced instrumental background \citep{acis_bkg} and astrophysical foreground and background components. For each ObsID, we generated an $0.5$--$7$ background-subtracted $S_X$ map from the corresponding clean $S_X$ map and blank-sky background event file with \texttt{blanksky$\_$image}. We merged the background-subtracted images for each ObsID to generate a single ZwCl 3146 clean background-subtracted $0.5$--$7$ counts map. 

Then, we calculated the X-ray peak of ZwCl 3146 from the clean background-subtracted $0.5$--$7$ counts map. We defined a set of circular annuli relative to this peak within the radial range of $25 < r < 625$ kpc, beginning with the innermost annulus and defining bin boundaries each time a threshold of $t_c = 7.5 \times 10^3$ background-subtracted counts was obtained. In this way, we ensure that each of the 11 annular bins will contain sufficient cluster counts to robustly obtain density and temperature values from a spectral fit. We then extracted source and background spectra for each ObsID in each annular bin. We defined an additional thin (20 pixel; $\simeq86$ kpc) outermost bin, which is used to initialize the spectral deprojection procedure, but not included in the pressure profile evaluation. With these spectra, we performed a spectral deprojection with \texttt{dsdeproj} \citep{Sanders2007,Russell2008} for each ObsID, assuming a spherical underlying source geometry beginning with our additionally defined outermost annular bin. When calculating the emission volumes in this deprojection, we included only the fraction of each annulus not contaminated by point sources. The result of this deprojection is a set of background-subtracted deprojected X-ray spectra for each annular bin. 

\section{Pressure profile generation} \label{sec:pressuregen}
\subsection{$P_{SZ}$} \label{subsec:SZpressure}
To derive the deprojected pressure profile in \citet{Romero2020}, annular surface brightness rings were fit to the MUSTANG-2 data, along with Gaussians to model the compact point sources, including the central AGN. In \citet{Romero2023}, another model component of a parabola in (telescope) elevation was added to further reduce large-scale noise. In principle, a reduction in the recovered source signal at large angular scales is possible from such an elevation correction, but any such reduction is expected to be small compared to the measurement uncertainties of the derived pressure profiles (see \citealt{Romero2023}, their Figure 3). The pressure model, which was then fit to the surface brightness profile, assumes a power-law pressure distribution within radial bins which are always spaced at least 1 beam-width apart \citep{Romero2020}. 

We begin with the MCMC chains for the pressure profile fits from \citet{Romero2023}. For each step, we evaluated the pressure profile at the X-ray defined bin centers (see Section \ref{subsec:X-raydata}). The outermost of these bins is evaluated at $\simeq1.64'$, which is well inside the $\simeq2'$ radius within which MUSTANG-2 reliably recovers signal in a minimally biased manner. The SZ-derived pressure profiles are centered at the position of the \textit{XMM-Newton} X-ray centroid, which is within an \textit{XMM-Newton} half-power diameter ($\simeq 15''$) of the more precisely determined \textit{Chandra} X-ray peak. \citet{Romero2020} note that the choice of centroid (derived from \textit{XMM-Newton} or MUSTANG-2, which are also separated within an \textit{XMM-Newton} PSF width) has a negligible effect on the derived pressure profile. Taking this, combined with the much larger PSF size of MUSTANG-2 ($\simeq 10''$ FWHM) relative to \textit{Chandra}, we therefore expect the offset between the X-ray- and SZ-derived pressure profile centers to result in negligible bias when comparing pressure values. From these realizations of the pressure profile, we estimated the median and $1\sigma$ uncertainty of $P_{SZ}$ in each bin, which comprise our $P_{SZ}$ profile (see Fig. \ref{fig:pressures}).

\subsection{$P_X$} \label{subsec:X-raypressure}
We performed a simultaneous fit of the background-subtracted deprojected spectra with \textit{Sherpa} for each ObsID in a given annular bin over an energy range of $0.5$--$7$ keV. We model the spectra as a single collisionally ionized plasma modified by interstellar absorption \citep[\texttt{tbabs $\times$ apec};][]{tbabs,Smith2001,Foster2012} with fixed hydrogen column $n_{\text{H}} = 2 \times 10^{20}$ cm$^{-2}$ \citep{absmap}, redshift $z = 0.291$ \citep[][]{Allen1992}, and metallicity $Z = 0.3\; Z_{\odot}$ with abundances from \citet{angrabunds}. The free parameters are thus the plasma temperature and normalization (which is itself a function of the electron and ion densities). 

We calculate the pressure from each spectral fit as the product of the electron number density $n_e$ and temperature $kT$. We derive $n_e$ from the fitted \texttt{apec} normalization: 
\begin{equation} \label{eq:norm}
    \eta = \frac{10^{-14}}{4\pi (D_A (1+z))^2} \cdot \int{n_e n_H\;dV}
\end{equation} for $D_A$ being the angular diameter distance to the source, $n_H$ being the H number density, $V$ being the volume of the source region, and assuming all units are CGS. The deprojected spectra are normalized to unit volume, and we assume $n_e \simeq 1.2 n_H$ for a fully ionized solar abundance plasma, and that the density and temperature are uniform within each bin. Then, the pressure in each annular bin is simply $P_X = n_e \cdot kT$. We estimate the $1\sigma$ uncertainty on $P_X$ by sampling 1000 random realizations of $P_X$ from $n_e$ and $kT$ drawn from the calculated posterior distributions of the $\eta$ and $kT$ parameters (see Fig. \ref{fig:pressures}).

At large radii, CC cluster pressure profiles derived from X-rays and the SZ effect should agree well. However, X-ray-derived temperature (and therefore, pressure) uncertainties associated with effective area calibration uncertainties exist in modern X-ray observatories, including \textit{Chandra}. These calibration uncertainties have been examined in great detail, and are generally expected to occur at the $\gtrsim 10$\% level \citep{Schellenberger2015,Migkas2024}, in contrast to the percent-level calibration possible for SZ data \citep{Wan2021}. Therefore, we choose to calibrate the absolute normalization of the $P_X$ profile to the $P_{SZ}$ profile (generated as described in Section \ref{subsec:SZpressure}). To calculate this correction factor and calibrate the uncertainty thereof, we generate 1000 realizations of $P_{SZ}$ and $P_X$ by randomly sampling the posterior distributions of each profile. In each realization, we estimate the mean value of $\alpha_P \equiv P_{SZ} / P_X |_{r >100\text{kpc}}$ for all bins with minimum bin edges outside of 100 kpc (of which there are six). The pressure profiles at these radii are less likely to suffer from projection effects associated with environmental complexity induced by the central AGN, and they enable accounting for biases in the absolute calibration of the X-ray-derived profiles relative to those derived from SZ without (by design) washing out the possible signal (decrement) that we seek to test at inner radii. 

We find that $\alpha_P = 0.93 \pm 0.04$, or that the X-ray derived pressure profiles are on average biased $\simeq7$\% higher than those derived from the SZ effect. This value of  $\alpha_P$ is consistent with the expected level of calibration uncertainty and the value derived from a comparison of the SZ- and X-ray (via \textit{Chandra})-derived pressure profiles for a large sample of clusters in \citet{Wan2021} (their normalization value being $\simeq0.91$). We thus scale the entire $P_X$ profile by $\alpha_P$ to ensure agreement between the pressure profiles at large radii (see Figure \ref{fig:pressures}), though see Section \ref{sec:results} for further discussion of this.

\section{Results} \label{sec:results}

\begin{figure*}[th!]
    \centering 
        \includegraphics[width=0.6\textwidth]{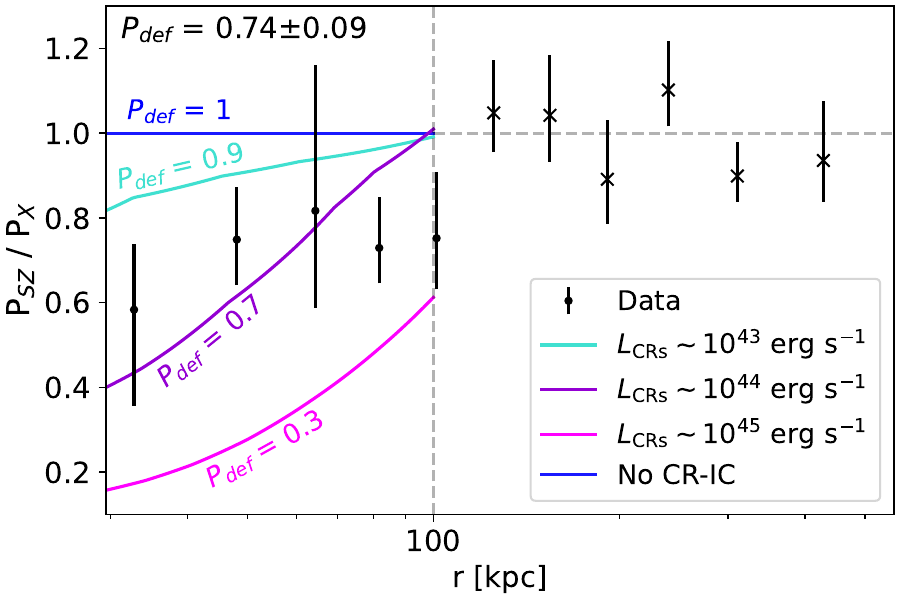} 
        \caption{Ratios of SZ-to-X-ray derived pressure profiles (black data points; $1\sigma$ uncertainties) with predictions from \citet{Hopkins2025CRs.CC} overplotted for various CR injection luminosities. The predicted deficit in $P_{\text{SZ}} / P_X$ at radii within $\simeq100$ kpc of the cluster center becomes more extreme for higher CR injection luminosities. Data points indicated with circles represent bins interior to 100 kpc used to calculate $P_{\rm def}$, while those indicated with crosses represent bins used to calculate $\alpha_P$. Our data from MUSTANG-2 and \textit{Chandra} indicate the presence of a deficit in  $P_{\text{SZ}} / P_X$ at $\simeq3\sigma$ significance.} \label{fig:ratios}
\end{figure*} 

We calculate the SZ-to-X-ray pressure deficit within 100 kpc of the X-ray peak from $P_{SZ}$ and $P_X$ by leveraging the 1000 realizations of $P_{SZ}$ and $P_X$ described in Section \ref{subsec:X-raypressure}. For each realization, we apply the mean value of $\alpha_P$ for that realization to $P_X$. Once $P_X$ is corrected, we estimate the (unweighted) mean ratio of $P_{SZ} / P_X$ for all bins with minimum bin edges within 100 kpc (of which there are five). We use this distribution of 1000 pressure deficit values to estimate the total SZ-to-X-ray pressure deficit: $P_{\text{def}} \equiv \langle P_{SZ} / P_X\rangle_{r\;\leq\;100\;{\rm kpc}}= 0.74 \pm 0.09$. While $P_{\text{def}}$ represents the average pressure deficit within 100 kpc, the values of $P_{SZ} / P_X$ within this radius consistently trend downwards as a function of decreasing cluster radius. 

This procedure encapsulates the variation in $\alpha_P$, which is not precisely constrained in the literature, within the statistical uncertainty estimate. In principle, the expected degree of X-ray calibration uncertainty is larger at higher plasma temperatures \citep{Schellenberger2015}, i.e., biased towards outer cluster radii for a CC. Even given the aforementioned motivation for normalizing $P_X$ using the outer radii values, we perform the analysis without applying this normalization correction, obtaining $P_{\text{def}} = 0.69 \pm 0.08$. Since the $\alpha_P$ correction shifts $P_X$ down relative to $P_{SZ}$, neglecting this correction (and its associated scatter) results in a downward shift of $P_{SZ} / P_X$, and thus results in a moderately more statistically significant measurement of $P_{\text{def}}$. 

In principle, our results are further dependent on two main modeling systematics; namely, the counts threshold used to define bin radii and the width of the outermost bin used in the spectral deprojection. To determine the importance of these two systematic uncertainties in our modeling, we evaluate a distribution of $P_{\text{def}}$ and $\alpha_P$ values using the above described procedures assuming a range of reasonable counts thresholds ($t_c = [6000, 6500, ..., 9000]$ counts) and outermost bin radii ($r_{w} = [10, 15, ..., 30]$ image pixels), for 35 combinations in total. The additional systematic uncertainty associated with our modeling procedures from these distributions is negligibly small relative to our statistical uncertainty: $\sigma_{P_{\text{def}}} \simeq 0.02$. We therefore report $P_{\text{def}}$ as above with the statistical uncertainty exclusively. 

Finally, while we evaluate the ZwCl 3146 $P_X$ profiles within a set of discrete annular bins assuming constant pressure within each bin, we calculate $P_{SZ}$ at a series of nodes corresponding to each linear bin center (where each node is connected to adjacent nodes via a power law). In principle, this difference could introduce a bias in the pressure profile comparisons. To test for this possible bias, we evaluate the $P_{SZ}$ profiles via two methods. First, as in the observational analysis, we estimate $P_{SZ}$ at each linear bin center. Second, we estimate $P_{SZ}$ at the ``X-ray emission-weighted center'', i.e., $\sqrt{\langle P^2\rangle}$ (assuming constant temperature within each bin). Within each bin, differences in $P_{SZ}$ evaluated via each of these procedures are far subdominant to the uncertainty on each respective pressure estimate, so we conclude that this methodology difference should not significantly bias our comparisons. 

\section{Discussion} \label{sec:discussion}

We detect a $\simeq 3\sigma$ deficit in the SZ-to-X-ray pressure ratio within 100 kpc of the ZwCl 3146 cluster center, $P_{\text{def}} = 0.74 \pm 0.09$. Although of moderate statistical significance, this result aligns with a scenario in which CR–IC emission non-negligably contributes to the X-ray continuum, artificially boosting the pressure inferred from X-rays. The models introduced in \citet{Hopkins2025CRs.gal,Hopkins2025CRs.CC} naturally predict this behavior, indicating that for realistic levels of AGN-driven CR injection, central CC pressure ratios can be suppressed to values as low as $P_{\text{def}} \simeq 0.3$ within 100 kpc. Of these, models assuming a CR injection luminosity $L_{\rm CRs} \sim 10^{44}$ erg s$^{-1}$ predict a central pressure deficit $P_{\text{def}} \simeq 0.7$ within 100 kpc, consistent with the ZwCl 3146 value. This demonstrates that a significant non-thermal X-ray continuum component could be present in ZwCl 3146 and plausibly generated by CR-IC in an ACRH. Energetically, a CR injection luminosity of order $L_{\rm CRs} \sim 10^{44}$ erg s$^{-1}$ is reasonable given the large observed jet kinetic power of the ZwCl 3146 AGN ($\sim6\times10^{45}$ erg s$^{-1}$; \citealt{Rafferty2006}), which is consistent with measured trends between the observed jet kinetic power and the X-ray luminosity \citep{Rafferty2006}. Such CR-IC would provide a natural physical mechanism for offsetting X-ray-inferred radiative cooling in CCs: by contributing $\sim$keV thermal continuum-like emission in the soft X-ray band, CR–IC scattering masks the true thermal budget of the ICM, thereby alleviating the apparent discrepancy of X-ray-inferred and true gas cooling rates at the heart of the cooling flow problem.

\subsection{Other possible systematics}\label{subsec:systems}
In this section, we explore additional possible systematics and physical processes that could contribute to the observed deficit in $P_{SZ} / P_X$. The most precise quantitative calibration of these effects would require a one-to-one comparison of full instrument mock SZ and X-ray pressure measurements derived from a cosmological sample of CCs, thus incorporating instrumental effects (e.g., PSF smearing, deprojection of a non-truncated gas distribution, etc.), and astrophysical effects (see below), which is beyond the scope of this work. Instead, we provide reasonable estimates for the expected contributions from the leading sources of bias in our analysis calculated via available analytic models or numerical calibrations. Given this, we do not attempt here to correct the measured $P_{\rm def}$ for these possible biases, but we provide quantitative estimates of these systematics for guidance in determining the plausibility of a CR-IC interpretation of the measured SZ-to-X-ray pressure deficit in ZwCl 3146. A summary of all contributions we consider and their relative importance to the measured SZ-to-X-ray pressure deficit in ZwCl 3146 is given in Table \ref{tab:contributions}. 

\subsubsection{Halo triaxiality + orientation} \label{subsec:elong} 
Dark matter halos are intrinsically triaxial, with shapes correlated with their formation histories \citep{Lau2021}. Within the gravitational potential of a cluster, the ICM inherits this triaxiality \citep{Machado2021}, and axial ratios of the ICM are expected to vary as a function of radius \citet{Lau2011}. Such asphericity affects both cluster selection (since elongated systems aligned with the LOS are preferentially detected; \citealt{Dietrich2014,Saxena2025}), and the interpretation of multiwavelength observables. Since X-ray and SZ observables depend differently on the LOS projection, neglecting gas triaxiality could introduce systematic biases in derived thermodynamic profiles \citep[e.g.,][]{Wan2021}.

We estimated the effects of halo triaxiality and orientation on our result as follows. We first obtained a tabulated distribution of average differential minor-to-major and intermediate-to-major ICM axial ratios as a function of cluster-centric radius derived from a sample of relaxed clusters in a $\Lambda$CDM cosmological simulation with gas cooling and star formation physics measured at low redshift \citep{Lau2011}. We then initialize a 3D grid with the axes of this average triaxial distribution aligned with those of the grid. We distribute a general (arbitrarily normalized) electron number density function over the triaxial distribution in the shape of a cuspy $\beta$-profile \citep{Vikhlinin2006}:

\begin{eqnarray} \label{eq:density}
    n_e \propto \frac{(r/r_c)^{-\alpha/2}}{((1 + (r/r_c)^2)^{3\beta/2 - \alpha/4}}
\end{eqnarray} for $\beta$-profile shape parameters typical of a CC cluster: the core radius $r_c = 0.1r_{2500}$, inner slope parameter $\alpha = [0.5, 1.5, 3]$ \citep{Bartalucci2023}, and $\beta$-profile parameter $\beta = 0.7$. For a single cluster of ZwCl 3146 mass, we obtain 1000 realizations of the triaxial halo viewed along randomly sampled line-of-sight (LOS) vectors. For this test, we note that we implicitly assume that a given cluster within a broader sample will be observationally selected agnostic to orientation, which, while not strictly true in general, is a reasonable approximation for most cluster samples selected from large surveys \citep[e.g.,][]{Saxena2025}. Then, to calculate the effects of the underlying triaxial gas density distribution on the SZ- and X-ray-derived pressure profiles, we create simple mock maps of the SZ and X-ray observables projected along the random LOS vectors assuming two cases: first, a triaxial gas distribution using the \citet{Lau2011} ICM axial ratio distribution (wherein $r$ in the above Equation \ref{eq:density} is the elliptical radius), and second, a spherical distribution (wherein $r$ in Equation \ref{eq:density} is the spherical radius and the axial ratios are set to 1). For a roughly isothermal distribution, the SZ-derived pressure is $P_e \propto \int n_e dl$ and the X-ray-derived pressure is $P_X \propto \sqrt{\int n_e^2 dl}$, so we project these quantities ($n_e$ and $n_e^2$, respectively) along the LOS vectors for each halo. 

We then performed a deprojection of these maps using an analogous spherical geometric deprojection formalism as the X-ray observational analysis (see Sec. \ref{subsec:X-raydata}). Since this deprojection routine implicitly assumes that the projected quantity is equal to zero outside the outermost deprojected radius, we apply a truncation to the density function at the outermost bin edge ($r_{outer} = 5$ Mpc) before the projection. Estimating $P_{\text{def}}$ in an analogous way as in the observational analysis for a finely spaced grid of bin edges sampled down to a radius of 10 kpc, we tested the effect of the choice of $\alpha$ using the spherical distributions. For (de-)projections generated at identical resolution as the observational data, we find no significant bias as a result of varying $\alpha$, though a systematic offset of $\simeq1-2\%$ can emerge for lower-resolution data when the deprojection is unable to recover the innermost shape of the density/pressure profiles given the steep gradient at small radii within each deprojection bin. Ultimately, we find the $1\sigma$ range of the recovered SZ-to-X-ray pressure ratio within 100 kpc for the radially-varying triaxial distributions across all tested $\alpha$ values is $P_{\text{def}} = 0.98_{-0.05}^{+0.06}$, or biased $\simeq2\%$ low on average with a few-percent-level  scatter. Thus the overall contribution to $P_{\text{def}}$ due to halo triaxiality and cluster orientation is likely insufficient to explain the measured ZwCl 3146 pressure deficit. 

\subsubsubsection{Higher-order ICM asymmetries}
Higher-order asymmetries in the ICM distribution beyond simple triaxiality are additionally possible, and indeed, the central region of the $S_X$ image of ZwCl 3146 appears to have such an asymmetry along the NW-SE axis (see Figure \ref{fig:pressures}). The lower S/N and less spatially-resolved SZ data are much less sensitive to such features, and do not indicate an analogous asymmetry. To determine whether this ZwCl 3146-specific asymmetry in $S_X$ is significant, we computed $P_X$ profiles extracted from two separate hemispheres along this axis. We find the ratio of $P_X^{\rm SE}$ to $P_X^{\rm NW}$ values within 100 kpc of the cluster center to be $1.13 \pm 0.09$, consistent with unity at the $1.5\sigma$ level. Thus, for ZwCl 3146, there is no statistically significant evidence for higher order asymmetries within the radial range relevant to our study given the measurement uncertainties associated with the existing data. More generally, to properly assess the potential impact of such asymmetries on the quantity of interest, $P_{\text{def}}$, a detailed analysis of $P_{\rm SZ}$ to $P_X$ profiles extracted from a large representative sample of CC systems from cosmological simulations is needed (see pt. 6 in Section \ref{sec:conclusions}). Such an analysis is beyond the scope of this work, and, as shown in our analysis of the hemispheric $P_X$ profiles, is unlikely to meaningfully impact our results for ZwCl 3146.

\begin{deluxetable*}{lll}[t]  
\caption{Census of possible contributions to the SZ-to-X-ray pressure deficit within 100 kpc of the ZwCl 3146 center. \label{tab:contributions}}
\tablehead{\colhead{Contribution} & \colhead{Expected effect} & \colhead{Additional considerations}}
\startdata
    \vspace{2mm} CR-IC & $1 \geq P_{\text{def}} \gtrsim 0.3$ & -- Tested across two orders-of-magnitude of CR injection luminosity \\ 
    \vspace{2mm}Halo triaxiality + orientation & $P_{\text{def}} = 0.98_{-0.05}^{+0.06}$ & -- Tested for a single model of radially-dependent halo triaxiality. \\  
    AGN-driven cavities & $P_{\text{def}} = 0.97_{-0.01}^{+0.01}$ & -- Tested for the pair of extreme observed cavities in ZwCl 3146 \\
    & & \textcolor{white}{-- }(nearly an order of magnitude larger in projected area than a  \\
    \vspace{2mm} & & \textcolor{white}{-- }typical AGN-driven cavity from \citealt{Rafferty2006}). \\  
    \vspace{2mm}Gas clumping & $P_{\text{def}} = 1.00_{-0.01}^{+0.01}$ & -- Limited sensitivity due to probing radii $\lesssim 0.5 R_{500}$ \\
    Helium sedimentation & $P_{\text{def}} = 1$ & -- Plasma instabilities, bulk motions and turbulence driven by AGN \\ 
    & & \textcolor{white}{-- }outflows and merger activity render such sedimentation to be \\
    \vspace{2mm}& & \textcolor{white}{-- }negligible \\ 
    Cosmological parameters & $P_{\text{def}} = 1$ & -- Normalization of $P_X$ to $P_{SZ}$ at large radii removes (constant in \\
    \vspace{2mm}& & \textcolor{white}{-- }radius) bias from $P_X \propto D_A(z)^{-1/2}$ \\
    X-ray calibration uncertainty & $P_{\text{def}} = 1$ & -- Bias of $\sim 20$\% possible in the absence of our large radius \\
    \vspace{2mm}& & \textcolor{white}{-- }normalization of $P_X$ to $P_{SZ}$ 
\enddata 
\tablenotetext{a}{Contributions to $P_{\text{def}}$ from CR-IC, and choice of cosmological parameters are quantified as a characteristic range derived from a discrete set of analytic models.}
\tablenotetext{b}{Contributions to $P_{\text{def}}$ from halo triaxiality and orientation, AGN-driven cavities, and gas clumping are quantified with a 1$\sigma$ range calculated via numerical modeling. } \vspace{-1.5em}
\end{deluxetable*} 

\subsubsection{AGN-driven cavities} \label{cavities}
Outbursts from central AGN are thought to drive jets of relativistic particles that can inflate bubbles in the surrounding ICM, carving out underdense regions (``cavities''; see, e.g., \citealt{Fabian2012.rev}), preferentially in CC clusters \citep{Prunier2025}. Such AGN/feedback-driven cavities have been observed as depressions in the X-ray surface brightness distributions in many clusters \citep[e.g.,][]{Birzan2004,Rafferty2006,HlavacekLarrondo2012}, and they are likely pressure-supported by relativistic cosmic rays \citep{Abdulla2019,Ehlert2019,Yang2019}. These underdensities of thermal plasma in the ICM, which have been observationally detected in ZwCl 3146 \citep{Rafferty2006}, could therefore represent a contaminant in our measured SZ-to-X-ray pressure profiles at small radii.

We have quantified the expected effect of AGN/feedback-driven cavities on $P_{\rm def}$ in the inner ICM as follows. We begin by distributing the same general (arbitrarily normalized) electron number density function over a spherical distribution in the shape of a cuspy $\beta$-profile as used above (Section \ref{subsec:elong}) when estimating the effects of halo triaxiality and orientation. To simulate the presence of cavities similar to those observed in ZwCl 3146, we first obtain the projected semi-major/minor axes ($a, b$) and projected radial distance from the cluster core to the cavity center for both ZwCl 3146 cavities measured by \citet{Rafferty2006}. We calculate the projected elliptical area of each cavity, and then approximate each ellipse as a circle with equal area to obtain the radius of a sphere with equal projected area as the measured ellipse. We then define $(x, y)$ in observer coordinates as the projection plane (of observation). For each cavity, we randomly select $(x, y)$ components that comprise the observed projected radial distance from the core to the cavity center. For each $\alpha$ value, we perform 100 realizations of the cavities by letting the $z$-component of the cavity position (in observer coordinates) be a randomly sampled value with magnitude $\leq 3 \times$ the maximum of the $x$ and $y$ components, since the projected $(x, y)$ distance is measured observationally \citep{Rafferty2006}, while the  $z$-component is not. Thus for each realization we obtain the $(x, y, z)$ position of the cavity center for a spherical representation of each ZwCl 3146 cavity with fixed approximations to the projected properties measured by \citet{Rafferty2006}. Then, we set the pressure proxy constructed from the electron number density distribution ($P_{SZ} \propto \int n_e dl$ and $P_X \propto \sqrt{\int n_e^2 dl}$) equal to zero within each cavity (assuming the pressure to be dominated by relativistic cosmic rays that do not contribute to either signal; \citealt{Hopkins2026_CC_details}). We evaluate the effects on $P_{\rm def}$ due to cavities in the spherical cases with and without cavities for all values of $\alpha$ considered in Section \ref{subsec:elong}, finding a $\simeq2-4\%$ decrement to the SZ-to-X-ray ratio within 100 kpc of the cluster center, with $P_{\text{def}} = 0.97_{-0.01}^{+0.01}$. On average, the presence of AGN-driven cavities similar to those observed in ZwCl 3146 thus can result in a bias in $P_{\text{def}}$ $\simeq3\%$ low on average with percent-level  scatter.

We further note that the ZwCl 3146 cavities are extreme relative to those measured in typical clusters. Per \citet{Rafferty2006}, the projected area of a cavity in ZwCl 3146 is $\simeq 3380$ kpc$^2$, located at a projected distance of $\simeq 50$ kpc from the cluster center. However, the median cavity across the full sample of \citet{Rafferty2006} has a projected area of $\simeq 380$ kpc$^2$ and a projected distance of $\simeq 20$ kpc from the cluster center, similar to recent predictions from the TNG-Cluster cosmological simulations \citep{Prunier2025}. Thus the ZwCl 3146 cavities are nearly an order of magnitude larger (in projected area) than a typical AGN-driven cavity from \citet{Rafferty2006}, and they are at larger radii which are sufficiently far from the cluster center that they fall into the typical radial range of our proposed CR-IC measurement technique. The cavity enthalpy of ZwCl 3146 is similarly exceptional \citep{Rafferty2006,Hlavacek-Larrondo2013.enthalpy}. For a more typical (median) cavity, the expected effect on the SZ-to-X-ray ratio via our methodology would be reduced to $\lesssim 1\%$. Thus, we expect the AGN-driven cavities in ZwCl 3146 (which are more extreme than typical) to be insufficient to explain our measured SZ-to-X-ray pressure deficit in the cluster.

\subsubsection{Helium sedimentation} \label{subsec:helium}
The X-ray emission from massive galaxy clusters is dominated by the thermal bremsstrahlung continuum generated via electrons free-free scattering with hydrogen and helium nuclei. Theoretical calculations have predicted that the larger helium nuclei can sediment into the gravitational potential of the cluster on timescales of order the cluster age \citep[e.g.,][]{Abramopoulos1981,Gilfanov1984}, resulting in an enhanced abundance of helium nuclei in cluster cores that could bias the electron pressure derived from X-ray emission if one were to assume an emitting ICM composed of primordial abundances. 

Several previous works have studied the effects of possible helium sedimentation on the X-ray observables of galaxy clusters \citep[e.g.,][]{Qin2000,Chuzhoy2003,Chuzhoy2004,Markevitch2007,Ettori2006,Peng2009,Bulbul2011}. In particular, the X-ray surface brightness dominated by thermal bremsstrahlung (including contributions from H and He ions) in a cluster may be written as
\begin{eqnarray}
    S_X & \propto \int_{LOS} (n_e n_H \Lambda_{e H} + n_e n_{He} \Lambda_{e He}) \;dl \nonumber \\
    & \propto n_H^2\; (1+2x)(1+4x)\; \Lambda_{eH}
\end{eqnarray}
for $x \equiv n_{He} / n_H$ such that $n_e = (1+2x)\; n_H$, with the X-ray band cooling function $\Lambda_{e(H/He)}$ describing the emission from the free-free scattering of electrons with hydrogen and helium nuclei, and neglecting continuum contributions from all elements heavier than helium \citep{Peng2009}. Then, the electron pressure scales as
\begin{eqnarray}
    P_e \propto \left(\frac{1+2x}{1+4x}\right)^{1/2}
\end{eqnarray}
for a fixed (measured) $S_X$, intrinsic gas temperature, and X-ray spectroscopic electron temperature. In reality, spectroscopic fits to obtain $n_e$ are dependent on the spectroscopically-fit gas temperature, which determines the shape of the thermal bremsstrahlung continuum, as well as influences the line emission, which is further dependent on the metallicity of the plasma, though the line emission is a very small contribution to the overall surface brightness. \citet{Ettori2006} note that the X-ray spectroscopically-fit temperature of a cluster does not vary significantly in the presence of helium enrichment. While the metallicity is dependent on the underlying helium abundance, we fix the metallicity to 0.3 $X_{\odot}$ in our analysis due to the large parameter degeneracy and limited constraining power of the metallicity parameter in our fits. For an ICM composed of primordial abundances ($X = 0.75$, $Y = 0.25$), we would have $x = 0.083$. Therefore, in the case where the He abundance were doubled from its primordial value, if we were to calculate the electron pressure measured at fixed $S_X$ assuming $x = 0.083$, we would overestimate the X-ray-inferred $P_e$ by $\simeq5\%$. Analogously, in the limiting case where helium dominates the mass of the ICM ($Y \to 1$), the most extreme SZ-to-X-ray deficit that could be generated due to the overestimate of the X-ray-inferred $P_e$ from assuming primordial abundances is $\simeq 0.76$, which, while similar to our measured deficit, is physically implausible. 

Nevertheless, using the prediction for the helium-to-hydrogen mass fraction radial distribution from the \citet{Peng2009} 1D helium sedimentation model applied to a cluster with ages of $t_{age}$ = 9 Gyr and $t_{age}$ = 11 Gyr with an X-ray spectroscopic temperature of $10$ keV \citep[distributed according to the profile of][]{Vikhlinin2006}, we calculate the overestimation of the electron pressure that would be inferred from an X-ray measurement as a function of cluster radius. Based on this simple laminar model, we estimate an equivalent SZ-to-X-ray pressure deficit within 100 kpc for the oldest evaluated cluster age ($t_{age}$ = 11 Gyr) of $P_{\text{def}} \simeq 0.9$. This (unrealistically) simple estimate is much less extreme than the measured deficit ($P_{\text{def}} = 0.74 \pm 0.09$), and contributions become even less important for lower cluster ages. 

However, 1D models like that of \citet{Peng2009} are likely an overly idealized estimate of the effects of He sedimentation in cluster cores. First, they do not include the effects of cluster environments, which are subject to, e.g., bulk motions or turbulence driven by AGN outflows, merger activity, etc. \citep{Zhuravleva2014,Lau2017}. To first order, we can scale the amount of He sedimentation expected in a simple laminar system to account for the effects of bulk mixing and turbulence like 

\begin{equation} \label{eq:he_analy}
    \eta_{\rm sed, mix} \sim \eta_{\rm sed} \cdot e^{-t_{\rm sed} / t_{\rm mix}}
\end{equation} with $\eta_{\rm sed}$ being the amount of He sedimentation predicted for a laminar system (here causing $P_{\text{def}} \simeq 0.9$) with the He sedimentation time $t_{\rm sed}$ and the mixing time $t_{\rm mix}$. Following \citet{Peng2009}, and assuming the cluster is approximately in hydrostatic equilibrium, we can write $t_{\rm sed} / t_{\rm mix}$ as a function of cluster-centric radius $R$

\begin{eqnarray}
    t_{\rm sed} / t_{\rm mix} \sim 12\: \left(\frac{\mathcal{M}_s}{0.1}\right) \cdot \left(\frac{R}{100\:{\rm kpc}}\right) \cdot \\
    \nonumber  \left(\frac{\lambda_{\rm mix}}{(f_b/0.2)(n_{\rm gas} / 10^3\:{\rm cm^{-3}})(T/10\:{\rm keV})^{5/2}}\right)
\end{eqnarray} for the sonic Mach number of the bulk flows or turbulence $\mathcal{M}_s = v_{\rm mix} / c_s$ (with the largest-scale bulk mixing velocity $v_{\rm mix}$ and sound speed $c_s$), $\lambda_{\rm mix} \sim 1$ being the ratio of the coherence length of the largest-scale coherent flows (in/outflows, turbulence) to the cluster-centric radius in cluster CCs, the magnetic suppression factor $f_B \sim 0.2$ for tangled magnetic fields, and the cluster temperature $T\sim 5$ keV and density $n_{\rm gas} \sim 10^{-2}$ cm$^{-3}$ at a radius of $\sim100$ kpc in a characteristic CC \citep[e.g.,][]{Truong2024}. Finally, applying radial scalings of density and temperature ($n_{\rm gas} \propto 1/R$, $T \propto R^{0.6}$ from $10-100$ kpc; \citealt{Romero2020}), we have 

\begin{equation} \label{eq:det}
    t_{\rm sed} / t_{\rm mix} \sim 6\: \left(\frac{R}{100\:{\rm kpc}}\right)^{1/2} \cdot \left(\frac{v_{\rm mix}}{100\:{\rm km\;s^{-1}}}\right)
\end{equation}

Assuming a characteristic turbulent velocity dispersion derived from observations of the Perseus CC with \textit{Hitomi} \citep[$\sim200$km s$^{-1}$;][]{Hitomi2016_perseus}, the effects of He sedimentation (via Equations \ref{eq:he_analy} and  \ref{eq:det}) would be reduced to $<1\%$ of those predicted in the equivalent laminar system at a radius of 100 kpc. Even this relatively quiescent turbulence level thus acts to significantly suppress He sedimentation. Simulations of Perseus-like clusters from TNG-Cluster support this observed level of velocity dispersion inferred from X-rays, and additionally suggest the coexistence of bulk flows in excess of $\sim 1000$ km s$^{-1}$ \citep{Truong2024}, which would significantly boost the suppression of He sedimentation, further reducing its effects.

Second, and more dramatically, the 1D simulations further neglect effects from small-scale plasma instabilities in the presence of a He composition gradient. These instabilities--primarily the heat-flux-driven buoyancy instability (HBI) at small cluster-centric radii and the magnetothermal instability (MTI) at large radii--would effectively erase the composition gradients predicted by the 1D simulations \citep[see][]{Berlok2015,Berlok2016}. In particular, a He composition gradient would act to increase the (already fast-growing relative to the sedimentation timescale) growth rates of such instabilities in cluster core regions, further re-distributing the ICM. Taking this in addition to the effects of turbulent/outflow-driven mixing on the He composition gradient outlined above, we conclude that helium sedimentation should have negligible effects on the observed pressure deficit in ZwCl 3146. 

\subsubsection{Gas clumping} \label{subsec:clumping}
Another possible systematic that could affect our pressure measurements is enhanced X-ray emission from the inhomogeneous density field (``gas clumping'') from ongoing mass assembly in cluster outskirts ($R \gtrsim R_{500}$; e.g., \citealt{Eckert2015}). While both $P_X$ and $P_{SZ}$ are linearly dependent on the electron temperature, the dependence on the electron density is quadratic in the case of $P_X$ (being derived spectroscopically via the deprojected $S_X$) and linear in the case of $P_{SZ}$ (being derived simply from the deprojected LOS integral of the electron pressure). Thus, any dominant bias in the ratio of $P_{SZ} / P_X$ as a result of gas clumping would enter via the X-ray emission. Both simulations \citep[e.g.,][]{Nagai2011,Battaglia2015,Planelles2017} and observations \citep[e.g.,][]{Eckert2015,Simionescu2011,Urban2014} have shown that the clumping factor (which can be interpreted as the ratio of mean to median deprojected $S_X$; see \citealt{Eckert2015}) in relaxed clusters is reasonably flat at the radial range within which we normalize $P_X$ to $P_{SZ}$ ($\sim0.1-0.4\;R_{500}$), and is expected to be negligible at smaller radii where we calculate $P_{\text{def}}$. 

Nevertheless, to quantify the effects of gas clumping on the measured ZwCl 3146 SZ-to-X-ray pressure deficit, we leverage the analytic form for the (square root of the) observed clumping factor ($\sqrt{C} (r)$) fit to 31 clusters observed with ROSAT/PSPC \citep{Eckert2015}. We assume ``true'' initial underlying distributions for $P_X$ and $P_{SZ}$ equal to unity, and we generate 1000 realizations of $P_X \cdot\sqrt{C} (r)$. Following the procedure outlined above for ZwCl 3146, we normalize $P_X \cdot\sqrt{C} (r)$ to $P_{SZ}$ using all bins outside of 100 kpc, and we estimate the resulting value of $P_{\text{def}}$ within 100 kpc. We find that including this gas clumping effect results in no statistically significant bias to $P_{\text{def}}$, with a $1\sigma$ range of $P_{\text{def}} = 1.00_{-0.01}^{+0.01}$, suggesting negligible bias and percent-level scatter. Given that the clumping factor across our cluster radial range is reasonably flat \citep{Eckert2015}, combined with our normalization of $P_X$ to $P_{SZ}$ outside of 100 kpc, the mild scatter and lack of a statistically significant bias in $P_{\text{def}}$ is expected and should not contribute significantly to the measured ZwCl 3146 pressure deficit. 

\subsubsection{Cosmological parameters} \label{subsec:cosmology}
The choice of cosmological parameters within a given flat $\Lambda$CDM model can affect the X-ray- and SZ-derived pressure values. While the conversion of angular to physical scale in the source frame to determine the proper radii at which the pressure profiles are evaluated is identical within both the SZ and X-ray analyses, the conversion of X-ray emissivity to electron density (via the \texttt{apec} normalization, see Eq. \ref{eq:norm}) scales as $D_A(z)^{-1/2}$ (when one factors in the cosmological dependence of the source region volume $V \propto D_A(z)^3$). Thus, assuming the most recent Planck cosmological parameterization \citep[$H_0 = 67.66$ km s$^{-1}$ Mpc$^{-1}$, $\Omega_{m, 0} = 0.31$;][]{Planck18} in place of the concordance cosmology would result in a $\simeq1.6\%$ change in $P_X \propto D_A(z)^{-1/2}$, independent of the radius at which $P_X$ is evaluated. However, in practice, our results are entirely unaffected by this bias, since we calibrate for this systematic via the normalization of $P_X$ to $P_{SZ}$ at large radii (see Section \ref{subsec:X-raypressure}). 

\subsubsection{X-ray instrument calibration} \label{subsec:xcal}
As stated in Section \ref{subsec:X-raypressure}, the measured ICM electron temperature (and therefore, pressure) can vary significantly across modern X-ray instruments, mostly likely due to effective area calibration uncertainties. For example, if \textit{Chandra} were to measure a plasma temperature characteristic of ZwCl 3146 ($\simeq7$~keV) in a standard band ($0.7-7$~keV), detectors on \textit{XMM-Newton} would measure a temperature $\simeq20\%$ lower \citep{Schellenberger2015}, and \textit{eROSITA} would measure a temperature nearly $\simeq40\%$ lower \citep{Migkas2024}. Given these large uncertainties and the lack of clarity regarding which temperature measurement is closest to the true underlying electron temperature, we normalize $P_X$ to $P_{SZ}$ using the radial bins outside of 100 kpc to eliminate this potential systematic bias (see again Section \ref{subsec:X-raypressure}). 

\subsubsection{AGN contamination} \label{subsec:agncont}
In principle, bright mm-wave emission from both the central AGN core and associated extended emission from its radio jets or lobes could bias the measurement of $P_{SZ}$ in the innermost radial bins, although in practice the steep-spectrum lobes are generally very dim at mm wavelengths and thus not relevant. By design, we have selected ZwCl 3146 in part due to the relatively low mm-wave brightness of its central AGN core \citep{Romero2020}. The central AGN in ZwCl 3146 is modeled as a compact point-like source \citep{Romero2020,Romero2023} and jointly fit with the pressure profile so that any uncertainty associated with the central AGN is captured in the fitting methodology. Given the half width at half maximum of the MUSTANG-2 PSF of $\simeq5''$ ($\simeq20$~kpc), the innermost $P_{SZ}$ radial bin ($\simeq33$~kpc) is evaluated well outside the anticipated radius of significant PSF smearing of the central AGN (and even more so for the X-ray observations, given \textit{Chandra's} PSF half-power radius of $\simeq0.4''$). In addition, because of the low surface brightness and steep spectral slope of the ZwCl 3146 radio minihalo, we do not expect it to contribute significantly at the mm-wave frequencies of interest \citep{Giacintucci2014}. Therefore, we do not expect any significant systematic bias associated with mm-wave emission from the ZwCl 3146 central AGN.

\begin{figure*}[t!]
    \begin{subfigure}[T]{0.49\textwidth}
        \centering
        \includegraphics[width=1\textwidth]{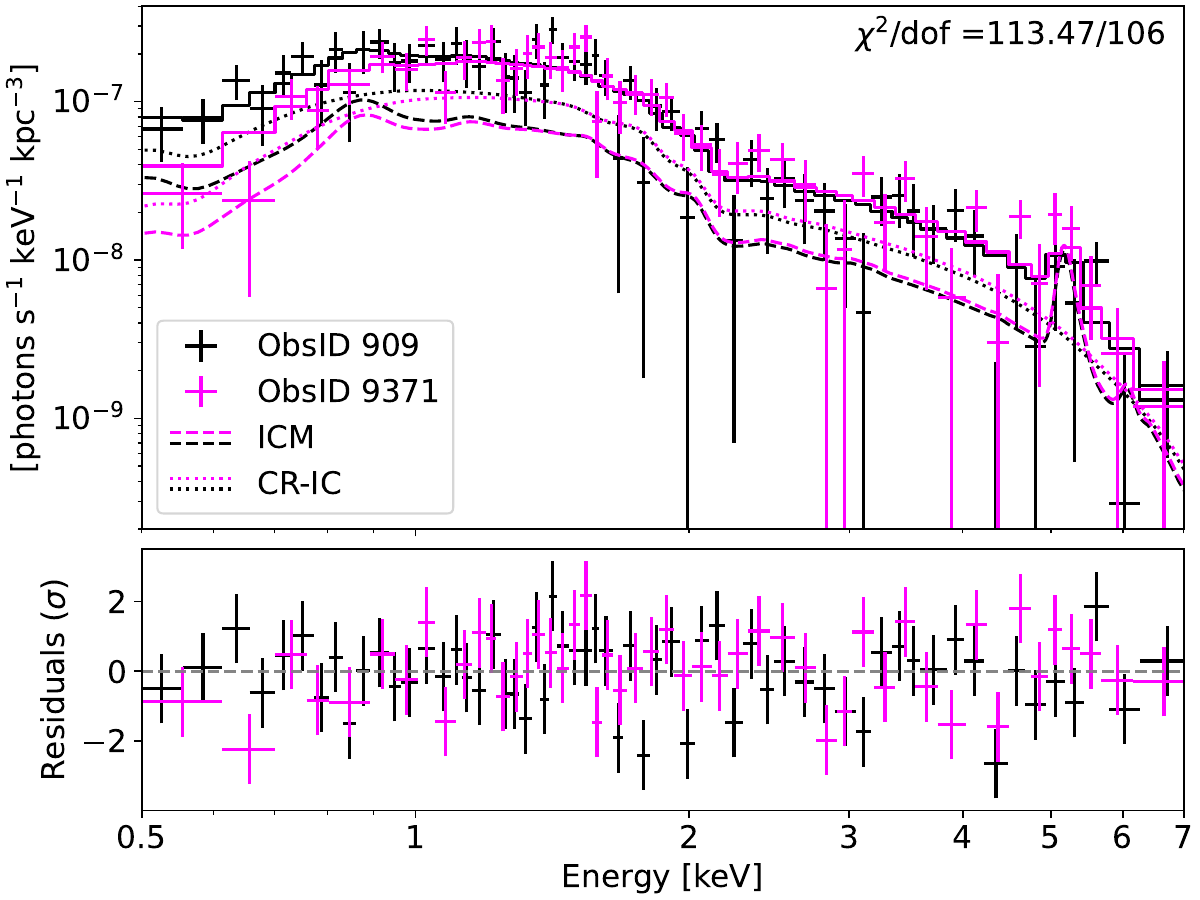}  
    \end{subfigure} \hfill
    \begin{subfigure}[T]{0.49\textwidth}
        \centering
        \includegraphics[width=1\textwidth]{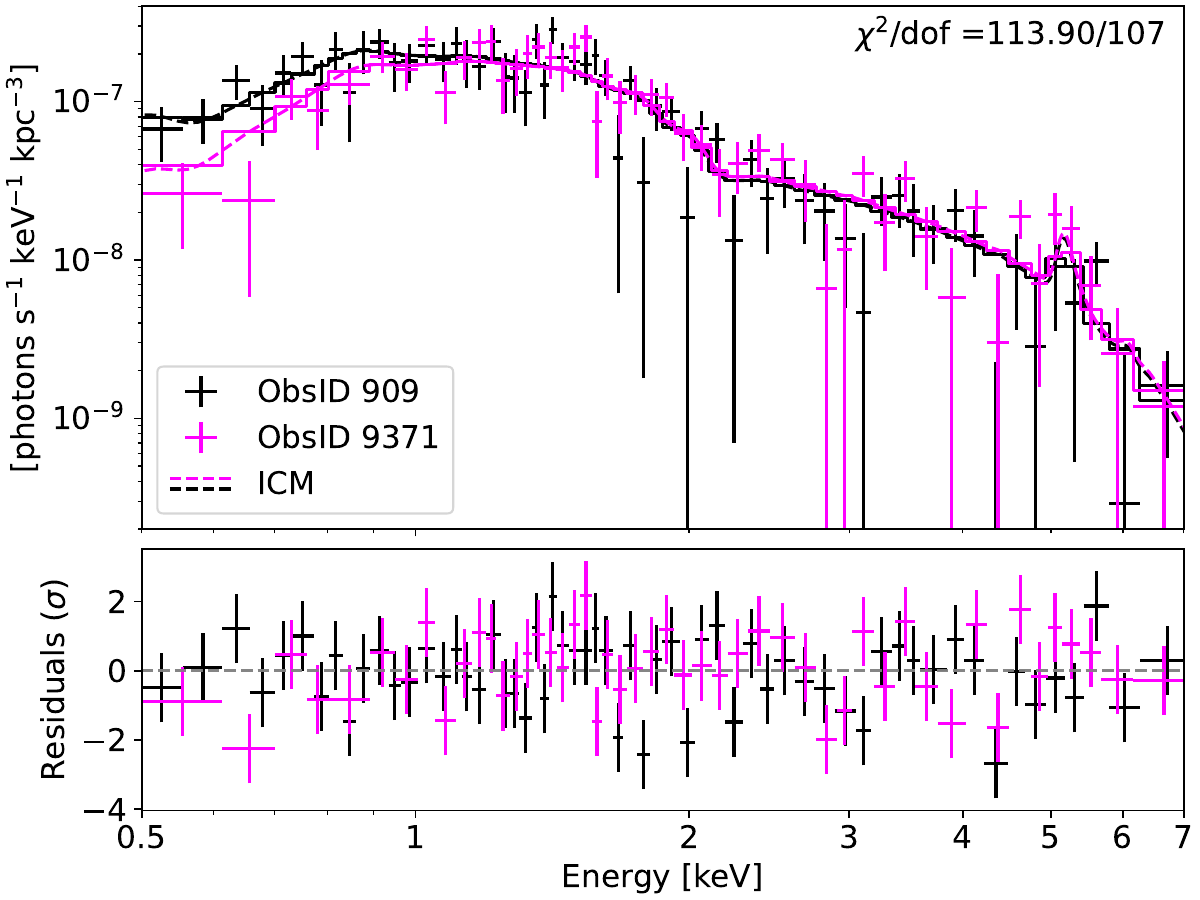}
    \end{subfigure}
        \caption{\textit{Left:} spectral fit of the innermost annular bin in the ZwCl 3146 core ($\simeq33$ kpc), with contributions from the fitted \texttt{tbabs $\times$ (apec + nlapec)} model for ICM thermal and CR-IC emission indicated. \textit{Right:} same, for the fitted \texttt{tbabs $\times$ apec} model for ICM thermal emission exclusively (no CR-IC component). This check confirms that, in the simple case where the CR-IC has a continuum shape similar to that of the thermal ICM continuum and a normalization determined by the measured SZ-to-X-ray pressure deficit in the innermost bin, the required metallicity of the thermal ICM ($Z_{\rm th} = 0.99 \pm 0.34\; Z_\odot$) is not unreasonably high, and the fit quality is relatively unchanged.} \label{fig:spec_CRIC}
\end{figure*}

\subsection{Fitting a CR-IC model to the X-ray spectra}\label{subsec:xspec}

Our measurement of an SZ-to-X-ray pressure deficit in ZwCl 3146 implies that a spectral component associated with CR-IC should be present in the inner core regions of ZwCl 3146. The measured ratio $P_{SZ} / P_X$ should reflect the ratio of the SZ-inferred true thermal pressure ($P_{SZ} \equiv P_{\rm th}$) to the X-ray-inferred pressure in the presence of CR-IC ($P_X \equiv P_{\text{th+IC}}$). To first order, when fitting a single absorbed \texttt{apec} model to the X-ray spectra, we obtain $P_{\text{th+IC}} \sim \eta_{\rm th+IC}^{1/2} kT_{\text{th+IC}}$, for $ \eta_{\rm th+IC}$ and $kT_{\text{th+IC}}$ corresponding to the normalization and temperature of that model. In order to obtain $P_{\rm th}$ from the X-ray spectra in the presence of non-negligible CR-IC, a two-component fit that includes the standard absorbed \texttt{apec} model plus a model of the CR-IC must be performed. In this scenario, $P_{\rm th} \sim \eta_{\text{th}}^{1/2} kT_{\text{th}}$ for $ \eta_{\text{th}}$ and $kT_{\text{th}}$ being the normalization and temperature of the absorbed \texttt{apec} model component. To assess whether the ZwCl 3146 X-ray spectra from the innermost radial bin ($\simeq33$~kpc, where a deficit in $P_{SZ} / P_X$ is detected at the highest statistical significance and where we thus expect the largest amount of CR-IC contamination) allow for the possibility of this CR-IC emission, we re-fit these data using a two-component model described above, with the following assumptions/simplifications. 

Performing such a fit requires a spectral model of the CR-IC emission, which is predicted to be similar in shape to thermal continuum emission. In particular, \citet{Hopkins2026_CC_details} predict that, as the CR electrons age and undergo Coulomb, IC, and bremsstrahlung losses--and with relatively weak dependence on the CR injection spectrum--the soft X-ray CR-IC spectrum will begin to exhibit curvature. This CR-IC spectral shape should mimic increasingly soft thermal bremsstrahlung continuum shapes as the CRs propagate out from the AGN injection site. Model calculations assuming an effective streaming speed $\simeq100$ km s$^{-1}$ \citep{Hopkins2026_CC_details} predict that after the CRs have propagated out to $r \sim 30$ kpc, the CR-IC spectrum will exhibit a shape similar to thermal bremsstrahlung continuum with an effective temperature of a $kT_{\text{IC}} \simeq 2.6$ keV. However, within the range of plausible CR injection spectra and CR propagation models, the effective temperature of the CR-IC emission could be as large as $kT_{\text{IC}} \simeq 10$~keV.

Given this large uncertainty, combined with the lack of constraining power of the X-ray data to simultaneously fit all of the CR-IC model parameters (often largely degenerate with the thermal ICM model parameters), we approximate the CR-IC spectral contributions for the simplest case: $kT_{\rm IC} = kT_{\rm th}$, and link these parameters in the spectral fit using the full model \texttt{tbabs $\times$ (apec $+$ nlapec)}, where the \texttt{nlapec} component represents our approximation of the CR-IC spectral component. In addition, we link the normalization of the CR-IC component to that of the thermal ICM component, applying a multiplicative factor which ensures that $P_{\rm th} / P_{\text{th+IC}}$ is equal to the measured SZ-to-X-ray pressure ratio for the innermost bin ($P_{SZ} / P_X = 0.58^{+0.16}_{-0.23}$), i.e., $\eta_{\rm IC} / \eta_{\rm th} = (P_{\rm SZ} / P_X )^{-2} - 1$ under the assumption that the continuum shape (thermal ICM temperature) remains constant. As in the observational analysis (see Sec. \ref{subsec:X-raypressure}), we fix the hydrogen column and redshift, and we leave the temperature and normalization of the thermal ICM component free. Since an additional continuum component should affect the X-ray-inferred metallicity of the plasma, we free the metallicity for the thermal ICM emission. We emphasize that this test leverages a rudimentary approximation of the expected CR-IC spectral shape \citep[motivated by the calculations presented in][]{Hopkins2025CRs.gal,Hopkins2025CRs.CC,Hopkins2026_CC_details}, and it is designed as an illustrative consistency check to determine whether a CR-IC continuum component at the level implied by our $P_{SZ} / P_X$ measurement is reasonably allowed by the data. 

The results of this two-component spectral fit compared to the single-component fit without CR-IC are shown in Figure \ref{fig:spec_CRIC} for the innermost ZwCl 3146 annular bin. When a CR-IC model with continuum shape approximated by $kT_{\rm IC} = kT_{\rm th}$ and a normalization linked to the thermal ICM component such that $P_{SZ} / P_X = 0.58$ is included, the fitted thermal ICM metallicity required to comprise the remaining emission from metal lines (predominantly Fe) is $Z_{\rm th} = 0.99 \pm 0.34\;  Z_\odot$. While a metallicity of $\sim Z_\odot$ is not unreasonably high at this distance from the cluster center and this measurement is consistent with the average Fe abundance profiles for clusters at this radius \citep{Mernier2017} within our relatively large measurement uncertainty and scatter in the average profiles, they are at the higher end of observed values. This is in general agreement with the picture outlined in \citet{Hopkins2026_CC_details}, where X-ray-inferred ICM metallicities derived without accounting for possible CR-IC continuum emission would be apparently suppressed relative to the ``true'' underlying metallicity for the same reason that the X-ray-inferred pressure is apparently boosted: the continuum is ``contaminated'' by CR-IC. There is a negligible difference in fit quality, and, as expected, the thermal ICM temperatures fitted in each case are identical ($kT = 4.30$ keV). In this simple case, the flux ratio in the 0.5--1.0~keV band for the CR-IC and thermal ICM components is $F^{0.5-1.0}_{\rm IC} / F^{0.5-7.0}_{\rm th} \simeq 1.4$. 

This test demonstrates that the innermost radial bin considered in our analysis of ZwCl 3146, near 33~kpc, allows for a substantial CR-IC continuum component with a spectral shape similar to the ICM thermal bremsstrahlung continuum with a normalization relative to the thermal ICM determined by our measured SZ-to-X-ray pressure deficit in the bin. However, this is a simple, illustrative consistency check, and we again stress that detailed modeling of such a CR-IC component should fully account for the uncertainties in the theoretical modeling, as well as marginalize over the possible systematics discussed in Section \ref{subsec:systems}. For example, given the broad range of theoretically motivated CR-IC spectral shapes, one should in principle fit for this shape using deeper X-ray observations. Additional theoretical modeling is also required for more complex cases where the CR-IC spectral shape is discrepant from the shape of the thermal ICM emission. In such a case, the addition of a softer/harder CR-IC spectral component will influence the X-ray-inferred ICM temperature relative to that derived from a single-component model fit. These temperature variations could then non-trivially influence the X-ray-inferred pressure values, and by extension, the ratio $P_{SZ} / P_X$. A study exploring these detailed models of CR-IC should further include the relevant radial information (e.g., $P_{SZ} / P_X$ trends as a function of radial bin) in a comprehensive fit. We leave such joint theoretical and observational exploration of these details for future work. 

\section{Summary \& conclusions} \label{sec:conclusions}
In this work, we have tested the possibility that CRs injected by a central AGN in CCs could non-negligibly bias the X-ray-inferred thermodynamic properties of such clusters by comparing X-ray and SZ observations of ZwCl 3146 from \textit{Chandra} and MUSTANG-2, respectively. Our main findings are summarized below: 

\begin{enumerate}
    \item We detect a $\simeq3\sigma$ (statistical) significance deficit in the SZ-derived pressure relative to that derived from X-rays ($P_{\text{def}} = 0.74 \pm 0.09$) within 100 kpc of the cluster center in ZwCl 3146. This decrement consistently trends downwards as a function of decreasing cluster radius at $r\lesssim100$~kpc. 

    \item This deficit is consistent with simple analytic models for CR injection by a central AGN introduced in \citet{Hopkins2025CRs.gal,Hopkins2025CRs.CC}, where $\sim$GeV CR leptons populate an ACRH within $\sim 100$ kpc of the central CC and IC scatter with CMB photons to produce thermal-continuum-like emission around $\sim$keV. The observed $P_{\text{def}} = 0.74$ is within the predicted range of models spanning two orders of magnitude in CR injection luminosity ($L_{\rm CRs} \sim 10^{43}- 10^{45}$ erg s$^{-1}$ \citealt{Hopkins2025CRs.CC}). This not only demonstrates that a significant non-thermal X-ray continuum component plausibly generated by CR-IC could be present in ZwCl 3146, but that such an effect could play a role in observed trends between AGN jet kinetic power and X-ray luminosity in CCs. 

    \item We estimated the effects of additional astrophysical, instrumental, and methodological contributions that could drive such an SZ-to-X-ray pressure deficit, including the effects of halo (gas) triaxiality and cluster orientation, AGN/feedback-driven cavities, helium sedimentation in the cluster potential, the ability of the deprojection routine to resolve the steep slope of the inner thermodynamic profiles, gas clumping due to mass assembly, choice of cosmological parameters, X-ray instrument calibration uncertainty, and mm-wave AGN contamination. In aggregate, these systematics can on average bias an $P_{SZ}/P_X$ pressure decrement measurement $\simeq5\%$ low, and the object-to-object variation in this systematic bias is $\simeq 5\%$ (see Table \ref{tab:contributions}). The leading-order systematics appear to be those associated with halo triaxiality/orientation and AGN-driven cavities. Systematics at this level are insufficient to fully explain our measured deficit of $\simeq 25$\% in ZwCl 3146, though more precise estimates would require robust forward-modeling of CC systems from cosmological simulations.  
    
    \item While the procedure for normalizing $P_X$ to $P_{SZ}$ at large radii ($> 100$ kpc) was designed to account for the leading-order systematic associated with the uncertainty of X-ray-derived temperature (pressure) measurements, we find that such a normalization mitigates biases associated with other astrophysical phenomena. Namely, the large-radius normalization of $P_X$ to $P_{SZ}$ reduces effects on $P_{\text{def}}$ due to halo triaxiality and cluster orientation, gas clumping, and choice of cosmological parameters. This normalization is thus a necessary step in comparing pressure profiles derived from X-rays and the SZ effect.    

    \item As a simple check, we fit the X-ray spectra of the innermost ZwCl 3146 radial bin ($\simeq33$ kpc, where the highest CR-IC flux is expected) with a two-component model including both CR-IC and thermal ICM, where the shape and normalization of the CR-IC component are provided by the thermal ICM component (under the assumption that the CR-IC component has a similar continuum shape as the thermal ICM continuum) and the measured SZ-to-X-ray pressure ratio, respectively. The X-ray spectra in this bin require a thermal ICM metallicity of $Z_{\rm th} = 0.99 \pm 0.34\; Z_\odot$ for a negligible change in fit quality. In this simple test case, the flux ratio $F^{0.5-1.0}_{\rm IC} / F^{0.5-7.0}_{\rm th} \simeq 1.4$.  

    \item We have shown that an SZ-to-X-ray pressure deficit within $\sim100$ kpc of a CC center can be observed at moderate statistical significance via measurements of the ZwCl 3146 system with existing instrumentation, and the systematics associated with such a pressure deficit can be straightforwardly handled. However, to make a robust claim of whether CR-IC contamination in the X-ray emission is contributing significantly to the cooling flow problem in CCs, more detailed studies accounting for the following are required: \textit{1.} additional observations of a representative population of CCs, \textit{2.} an empirical calibration of (instrumental, astrophysical, and methodological) systematics derived from mock observations of a large sample of CC systems in cosmological simulations, \textit{3.} an exploration of theoretical modeling choices/assumptions, including the effects of CR-IC spectral shapes discrepant from the thermal ICM spectral shapes, and \textit{4.} simultaneous handling of spectral and radial information in each cluster. 
\end{enumerate}

\begin{acknowledgments}
We thank the anonymous reviewer for their detailed comments, which helped to improve the presentation of our results. EMS thanks Sunil Golwala and Sam Ponnada for helpful discussions. EMS acknowledges support from a National Science Foundation Graduate Research Fellowship (NSF GRFP) under Grant No. DGE‐1745301. EMS and JS acknowledge the support from NASA/80NSSC25K0597. Support for PFH was provided by a Simons Investigator Award.
\end{acknowledgments}

\facilities{CXO, MUSTANG-2}

\software{CIAO \citep{ciao}, Xspec 12.12.0 \citep{arnaud1996}, Astropy \citep{astropy2013, astropy2018}, mpi4py \citep{dalcin2021}, NumPy \citep{vanderwalt2011, numpy}, Matplotlib \citep{matplotlib}}

\bibliography{refs}{}
\bibliographystyle{aasjournalv7}

\end{document}